# A Self-Supervised Learning of a Foundation Model for Analog Layout Design Automation

Sungyu Jeong†, *Graduate Student Member, IEEE*, Won Joon Choi†, *Graduate Student Member, IEEE*,
Junung Choi, *Graduate Student Member, IEEE*, Anik Biswas, *Graduate Student Member, IEEE*,
and Byungsub Kim, *Senior Member, IEEE*

*Abstract*—We propose a UNet-based foundation model and its self-supervised learning method to address two key challenges: 1) lack of qualified annotated analog layout data, and 2) excessive variety in analog layout design tasks. For self-supervised learning, we propose random patch sampling and random masking techniques automatically to obtain enough training data from a small unannotated layout dataset. The obtained data are greatly augmented, less biased, equally sized, and contain enough information for excessive varieties of qualified layout patterns. By pre-training with the obtained data, the proposed foundation model can learn implicit general knowledge on layout patterns so that it can be fine-tuned for various downstream layout tasks with small task-specific datasets. Fine-tuning provides an efficient and consolidated methodology for diverse downstream tasks, reducing the enormous human effort to develop a model per task separately. In experiments, the foundation model was pre-trained using 324,000 samples obtained from 6 silicon-proved manually designed analog circuits, then it was fine-tuned for the five example downstream tasks: generating contacts, vias, dummy fingers, N-wells, and metal routings. The fine-tuned models successfully performed these tasks for more than one thousand unseen layout inputs, generating DRC/LVS-clean layouts for 96.6% of samples. Compared with training the model from scratch for the metal routing task, fine-tuning required only 1/8 of the data to achieve the same dice score of 0.95. With the same data, fine-tuning achieved a 90% lower validation loss and a 40% higher benchmark score than training from scratch.

*Index Terms*—analog layout design automation, foundation model, self-supervised learning, fine-tuning, a consolidated methodology for various layout tasks

This work was supported in part by Institute of Information and Communications Technology Planning and Evaluation grant funded by the Korea Government (MSIT) (No. 2022-0-01171); in part by BK21 FOUR Project of NRF for the Department of Electrical Engineering, POSTECH; in part by Next-generation Intelligence semiconductor R&D Program through the National Research Foundation of Korea (NRF) funded by Korea government (MSIT) (RS-2023-00258227).

Sungyu Jeong, Won Joon Choi, Junung Choi, and Anik Biswas are with the Department of Electrical Engineering, Pohang University of Science and Technology, Pohang-si 37673, South Korea. (Sungyu Jeong and Won Joon Choi contributed equally to this work.)

Byungsub Kim is with the Department of Electrical Engineering, the Department of Convergence IT Engineering, and the Department of Semiconductor Engineering, Pohang University of Science and Technology, Pohang-si 37673, South Korea, and also with the Institute for Convergence Research and Education in Advanced Technology, Yonsei University, Seoul 03722, South Korea (e-mail: byungsub@postech.ac.kr).

## I. INTRODUCTION

FOUNDATION models [1] are game changers that are greatly impacting numerous applications from natural language processing [2] to visual synthesis [3] by providing revolutionary solutions for many problems in building, training, and deploying machine learning models. However, a foundation model for design automation of analog layout and its training method have not been studied yet.

For the first time, we propose a self-supervised learning method and a foundation model to address two major problems in applying deep learning to analog layout design automation: 1) lack of qualified annotated analog layout data, and 2) excessive variety in analog layout design tasks.

Lack of data poses a challenge in machine learning for analog layout [4]. Training requires lots of high-quality labeled analog layouts. Although flipping and rotation can augment data by a few times [4], it is insufficient for diverse tasks. Also, labeling requires lots of human labor. Automatic labeling can reduce human labor [4], [5], but developing algorithms for diverse tasks also burdens designers. Layout generators can produce many labeled analog layouts [6], but preparing generators for various circuits and ensuring the qualities of all generated layouts are also burdensome.

The proposed self-supervised learning resolves the problem of lack of annotated analog layout data for training. We obtained enough training data from unannotated manually-designed layouts by random patch sampling and random masking techniques.

The excessive variety of analog layout design tasks is also critical in machine learning for analog layout automation. So far, engineers must explicitly build each model for each of many diverse design tasks, spending too much time.

From unannotated data, the foundation model can be pre-trained implicitly for general knowledge on appropriate layout patterns, and then can be fine-tuned for various specific tasks to suggest the proper layout modifications based on the contexts of incomplete layout patterns as well as embedded commands. To the best of the authors' knowledge, our model is the first foundation model for analog layout design because its behavior is implicitly induced and it provides a consolidated methodology for various downstream tasks [1].

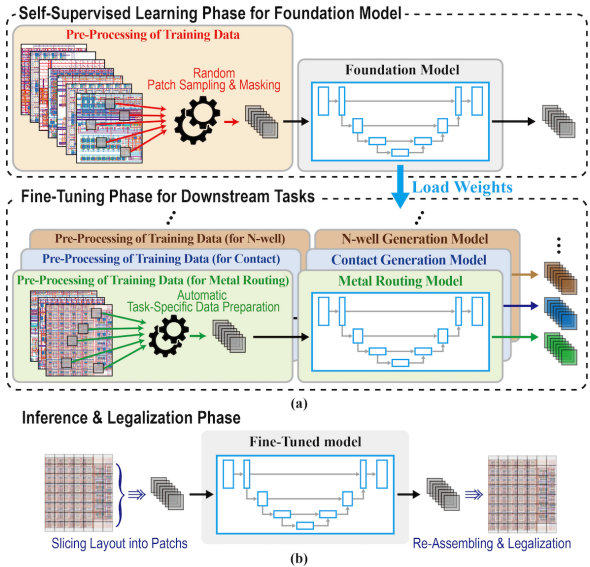

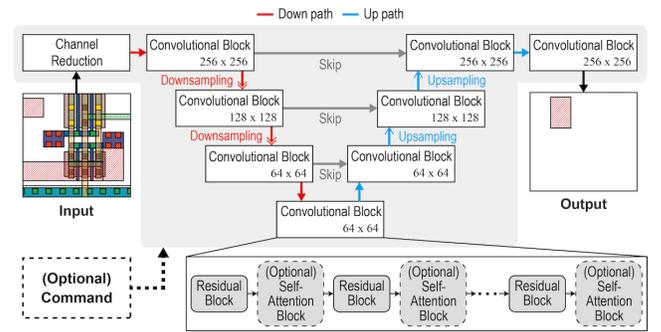

Fig. 2. Foundation model architecture.

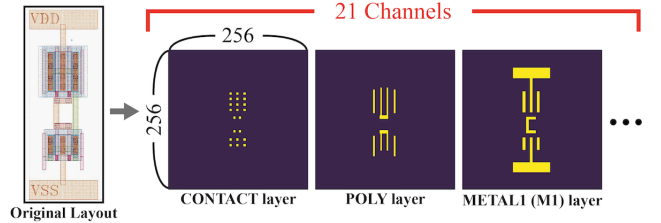

Fig. 1. Overview of the proposed workflow. (a) Self-supervised learning phase for foundation model training and fine-tuning phase for downstream tasks (b) Inference and legalization phase showing patch-based processing and layout reconstruction.

Fig. 3. Layout-to-patch conversion.

## II. PROPOSED METHOD

### A. Proposed Workflow

The proposed workflow has three phases: 1) a self-supervised learning phase, 2) a fine-tuning phase, and 3) an inference and legalization phase (Fig. 1).

In self-supervised learning, a foundation model is pre-trained using unannotated layout data (Fig. 1a). By randomly sampling layout patches and masking layout elements, the pre-training dataset is prepared without manual annotation. The dataset contains enough varieties of unbiased layout patterns with the same size and the same resolution so that the foundation model can implicitly learn general knowledge on proper layout patterns within a patch from these data.

During the fine-tuning phase, the pre-trained foundation model is quickly fine-tuned for specific downstream layout tasks using small amounts of task-specific data (Fig. 1a). Fine-tuning data for each downstream task are prepared by an algorithm that automatically selects and removes the target layout elements, similar to the self-supervised learning, but using a dedicated algorithm rather than random selection.

In the final inference and legalization phase, the fine-tuned model generates the desired layout elements based on the incomplete input layout pattern and the embedded command (Fig. 1b). The input layout is uniformly sliced into patches and then fed to the fine-tuned model. The output patches are reassembled to reconstruct the full layout result. This patch-based inference provides easy handling of layouts in various sizes while guaranteeing the same layout resolution as the model is trained with. At the end, the inferred layout result is legalized to comply with the design rules.

This workflow is applicable to diverse downstream layout automation tasks without task-specific model development.

### B. Model Architecture

We adopted UNet architecture [7] with 21 input channels representing 21 different physical layers of a layout design (e.g., metal, diffusion, via) (Fig. 2). Because UNet can preserve detailed geometric information and can provide precise pixel-level inference, it enables generation of high-quality layouts. To enhance inference capability on complex patterns, we increased the model's depth by incorporating residual convolutional blocks to mitigate the vanishing gradient problem [8]. Additionally, self-attention blocks were integrated to capture long-range dependencies, including relationships between distant layout elements [9].

The model's input is 256x256 pixel matrices of layout patches sampled from a layout design (Fig. 3). Unlike [5], which uses a 3-channel RGB image, we use a matrix with 21 channels corresponding to different physical layers. It is important that patches are sampled without resizing, covering a fixed layout area of 2,560 nm x 2,560 nm while each pixel size covers 10 nm x 10 nm. This patch format ensures that the model always deals with the data input of the same size and the same resolution regardless of the original layout size.

Similar to the UNet design [7], the model has a down path and an up path with skip connections (Fig. 2). To decrease computational cost, the down path reduces the channel counts of the input from 21 to 8. The down path then extracts multi-scale feature maps through residual blocks and downsampling layers enhanced by optional self-attention blocks. The up path reverses this process using upsampling. The skip connections aid accurate localization during reconstruction. This architecture enables the model to generate proper layout elements while preserving detailed spatial information.

Inspired by SAM [10] that uses prompts for image segmentation, explicit commands can be embedded and processed in our model along with layout patch data. The context of the layout pattern of an incomplete input layout patch does not always provide enough information to infer generation of the target layout elements. For example, multiple options for metal routing might be considered legitimate if inference relies solely on the layout patterns. In such circumstances, additional information such as the objective of the task can be provided by an embedded explicit command. For example, in metal routing, a matrix-formatted command indicates the start and the end points of routing to clarify the goal of the routing. This command is processed along with the patch data to infer generation of routing metals.

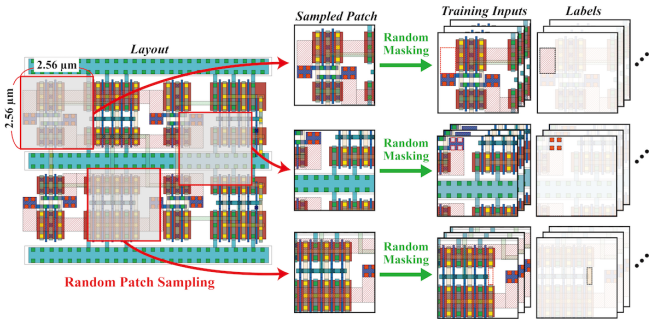

Fig. 4.   Illustrations of the proposed random-patch sampling and random masking techniques in self-supervised learning.

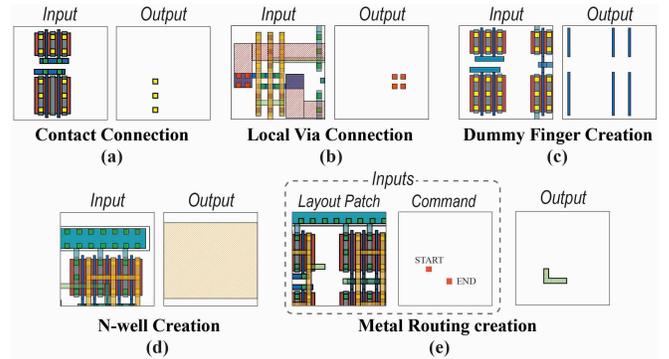

Fig. 5.   Example training data in fine-tuning for downstream tasks: (a) contact connection, (b) local via connection, (c) dummy finger creation, (d) N-well creation, and (e) metal routing creation with explicit command (indicating the start and end points of routing).

### C. Self-supervised Learning of the Foundation Model

A self-supervised learning method is enabled by two pre-processing techniques: 1) random patch sampling and 2) random masking (Fig. 4).

During random patch sampling, patches of the same size and the same resolution are sampled at random locations. Because layout data are greatly augmented by sampling many times and less biased by sampling at random locations, the dataset provides enough varieties of unbiased layout patterns as well as uniformity of size and resolution. Therefore, this method enhances the generalization ability of the foundation model.

During random masking, layout elements are randomly removed from patches. The remaining patches are used as the inputs for pre-training while the removed layout elements are used as the target outputs. This process greatly augments data because there are a lot of options for element removal. Moreover, we can automatically acquire enormous input-output pairs without manual labeling.

The foundation model is trained from unannotated layout data using the proposed self-supervised approach. By learning from diverse and unbiased patch data, the model implicitly acquires general knowledge on diverse appropriate layout patterns, and therefore, can easily adapt to various downstream tasks.

### D. Fine-Tuning of the Foundation Model

Fine-tuning provides an efficient and consolidated methodology to obtain models for diverse layout design tasks. Because the foundation model obtained general knowledge on the appropriate layout patterns, fine-tuning for a specific downstream task requires less training data and less computation than training models from scratch. Fine-tuned models can infer appropriate design modifications from incomplete layouts and commands, such as suggesting N-well regions for a given layout missing N-wells.

To prepare fine-tuning data, we use an automated process similar to the self-supervised learning method but modified for each task. The algorithm has two steps. 1) It samples patches from layout regions containing the target layout elements for a specific task, rather than randomly sampling. 2) The algorithm selectively removes the target layout elements from the sampled patches. For example, for N-well generation task, the algorithm samples patches from regions with N-well layers and removes all N-well layers from the sampled patches. This process creates input-output pairs of fine-tuning data, where the inputs and outputs are the layout patches without the target elements and the target layout elements themselves, respectively (Fig. 5).

In addition, a command can be prepared in a multi-channel matrix format in order to provide task-specific information. This matrix contains context information that is not inferable from the input patch, providing additional guidance to the model. For example, in the metal routing task, the command matrix indicates the start and end points of routing (Fig. 5e). This command is concatenated with the layout patch.

The foundation model is fine-tuned with the prepared fine-tuning data. The pre-trained weights of the foundation model are loaded to a new model while weights of any new layers for the specific task are initialized randomly. The new model is then trained using the prepared dataset with a low learning rate.

In summary, fine-tuning provides a consolidated methodology that is applicable to diverse analog layout tasks. By leveraging its general knowledge obtained in pre-training, the foundation model can easily and quickly adapt to specific downstream tasks using less training data and fewer computational resources than the model trained from scratch.

### E. Inference and Legalization

During inference, layouts larger than the patch size can be processed using a slicing and reassembling approach (Fig. 1b). We slice an input layout into uniformly-sized patches, ensuring that the model's input has the same resolution as in the training phase. Each patch is processed by the fine-tuned model and the inferred output patches are reassembled in the slicing order to obtain the full layout. Therefore, this approach

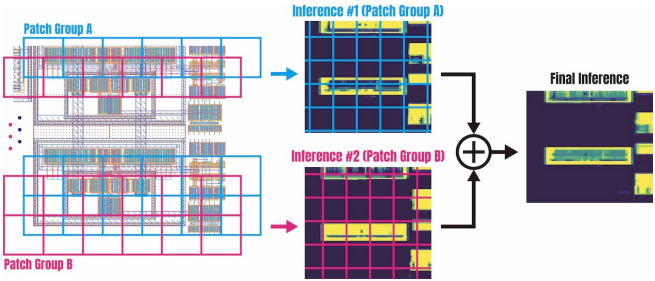

Fig. 6. Overview of the overlapping double patch sampling method. Final inference is derived by combining results from overlapping patch groups A and B on the original layout, with their respective inferences.

allows the model to work with large layouts in various sizes.

In the slicing and reassembling approach, the inference accuracy may degrade at patch borders because the model lacks context information beyond the patch boundaries.

To address this issue, we propose an overlapping double patch sampling method (Fig. 6). The input layout is differently sliced twice, creating two different patch groups. Every patch in one group overlaps with four patches from the other group to ensure that edge pixels in one group are interior pixels in the other group. Using this approach, we obtain two inference results per pixel. By combining the two results, we can resolve the inaccuracy issue at the patch borders.

Finally, we perform legalization on the inferred layout to ensure that the final layout complies with the design rules. The heuristic nature of deep learning may cause pixel-level inaccuracies resulting in not-clear boundaries of the inferred layout. Legalization including rectilinearization and polygon edge adjustment makes the final layout meet the design rules.

## III. Experimental Result

The foundation model was pre-trained on four A100 GPUs using training datasets obtained from manually designed layouts of 6 silicon-proved analog/mixed-signal circuits (Fig. 7 and 8). In the proposed self-supervised learning, we obtained about 15,000 patches in size of 256x256 by random patch sampling and then prepared a pre-training dataset of 324,000 patches from the sampled 15,000 patches by random masking. In order to make the dataset as unbiased as possible, we applied the random masking technique to each layer type an equal number of times. The pre-trained foundation model was fine-tuned for five tasks of generating (1) contacts, (2) vias (for M1-M2, M2-M3, and M3-M4 connections), (3) dummy fingers, (4) N-wells, and (5) metal routings (for M1, M2, and M3 layers) by using 50 patches, 1,500 patches, 1,100 patches, 1,700 patches, and 2,900 patches, respectively.

We prepared benchmark layouts that were not included in the training data in order to assess each model's inference performance for unseen layout inputs. Our benchmark includes various analog circuit types, some of which have multiple design variations. For each task, target layout elements were intentionally removed, and then the model's ability to accurately infer and to generate these missing elements was evaluated using DRC/LVS verifications of the result layouts. Table I lists the metrics used to evaluate the inference performance of each task.

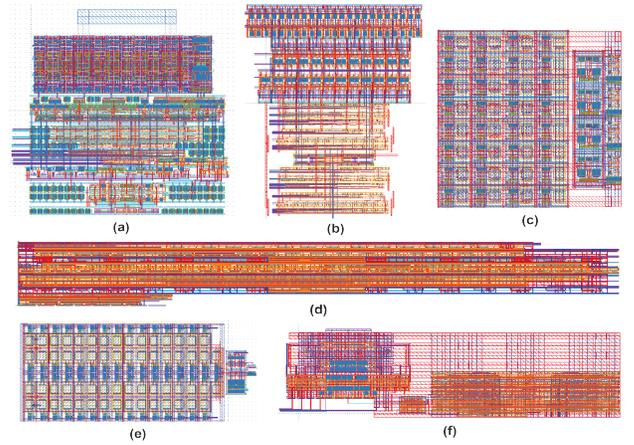

Fig. 7. Source layouts of silicon-proved, manually-designed analog/mixed-signal circuits used to extract training data. The layouts of (a)-(f) are differently scaled.

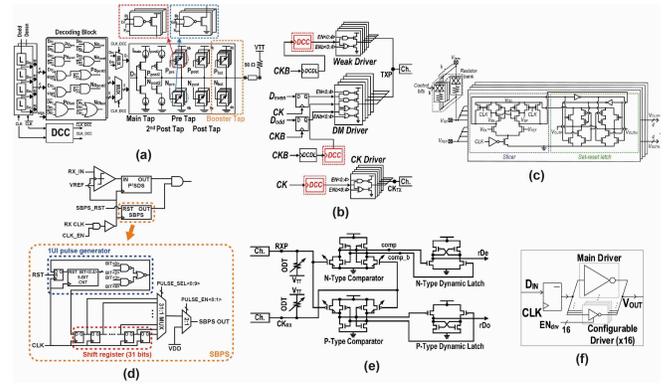

Fig. 8. Schematic diagrams of source layouts used for extracting training data, corresponding to the layouts in Fig. 7: (a) 20 Gb/s 4-tap feed-forward equalization TX; (b) 20 Gb/s data-embedded clock signaling (DECS) TX; (c) 20 Gb/s compact RX; (d) 8Gb/s sampling monitor; (e) 20 Gb/s DECS RX; (f) 50 Mb/s inverter-based TX. Circuits (a)-(e) are high-speed I/O circuits, while (f) is designed for body channel communication.

For each task of generating contacts, vias, dummy fingers, and N-wells, we prepared 300 benchmark layouts from 300 different source layout designs of three high-speed circuit types: wireline receivers (RXs), strong-arm sense amplifiers, and high-speed set-reset latches. For each circuit type, 100 different incomplete layout designs were prepared: for the task of generating contacts or vias, some contacts or vias were deleted, respectively, to cause LVS failure; for the generation of dummy fingers or N-wells, all dummy fingers or all N-wells were deleted, respectively.

For the metal routing generation task, we prepared 593 benchmark layouts from 28 different source layouts having 22 different circuit topologies. These circuits include CML-to-CMOS buffers, D Flip-Flops, sense amplifiers, high-speed MUXs, an operational amplifier (Op-Amp), high-speed set-reset latches, a true single phase clock (TSPC) latch, wireline transmit drivers (TX drivers), and a current summation circuit (Table III). From these layouts, we created benchmark samples by selectively deleting metal routing. Because the model for the metal routing task takes both a layout patch input and an explicit command input, we also prepared the command inputs that specify the starting and ending points of the routing tasks.

TABLE I.  DEFINITION OF THE METRICS USED TO ASSESS THE FINE-TUNED MODELS

| Tasks | Metrics |
|---|---|
| Contact or Local Via Connection | Success Ratio: <br> $\dfrac{\text{\# of correctly generated layouts in top-}k^{*}\text{ recommendations}}{\text{\# of benchmark data samples}}$ |
| Dummy Finger or N-well Creation | Intersection over Union (IoU) Score: <br> $\dfrac{\text{Area of (predicted region} \cap \text{ground truth region)}}{\text{Area of (predicted region} \cup \text{ground truth region)}}$ |
| Metal Routing Creation | Dice Score: <br> $\dfrac{2 \times \text{Area of (predicted region} \cap \text{ground truth region)}}{\text{Area of predicted region + Area of ground truth region}}$ |

*k=1 for contact, k=10 for local via

TABLE II.  NUMBERS OF SUCCESSFULLY GENERATED LAYOUTS FOR FOUR DOWNSTREAM TASKS: CONTACT, LOCAL VIA, DUMMY FINGER, AND N-WELL

| Cell Types | Contact | | Local Via | | Dummy finger | | N-well | |
|---|---|---|---|---|---|---|---|---|
| | DRC/LVS Passed | Benchmark Samples | DRC/LVS Passed | Benchmark Samples | DRC/LVS Passed | Benchmark Samples | DRC/LVS Passed | Benchmark Samples |
| Receiver | 100 | 100 | 100 | 100 | 100 | 100 | 98 | 100 |
| Sense Amp | 100 | 100 | 90 | 100 | 100 | 100 | 100 | 100 |
| Set-Reset Latch | 100 | 100 | 75 | 100 | 100 | 100 | 100 | 100 |

TABLE III.  NUMBERS OF SUCCESSFULLY GENERATED LAYOUTS FOR METAL ROUTING

| Cell Types | Straight Routing | | Bended Routing | | Notes |
|---|---|---|---|---|---|
| | DRC/LVS Passed | Benchmark Samples | DRC/LVS Passed | Benchmark Samples | |
| CML-to-CMOS Buffer1 | 31 | 31 | 6 | 8 | - |
| CML-to-CMOS Buffer2 | 23 | 23 | 6 | 6 | - |
| D Flip-Flop1 | 5 | 5 | 6 | 6 | - |
| D Flip-Flop2 | 10 | 10 | 10 | 10 | - |
| D Flip-Flop3 | 19 | 19 | 9 | 9 | - |
| D Flip-Flop4 | 35 | 35 | 41 | 42 | - |
| Double-Tail Sense Amp1 | 8 | 8 | 1 | 1 | N-type, with a latch |
| Double-Tail Sense Amp2 | 5 | 5 | 4 | 6 | P-type, with a latch |
| 2:1 MUX* | 19 | 19 | 5 | 5 | Transmisson-gate based |
| 4:1 MUX | 20 | 20 | 45 | 46 | For high-speed I/O |
| Operational Amp | 9 | 9 | 2 | 2 | - |
| Phase Detector | 35 | 35 | 28 | 30 | - |
| Strong-Arm Sense Amp1** | 34 | 34 | 24 | 25 | - |
| Strong-Arm Sense Amp2 | 19 | 19 | 7 | 14 | With threshold control |
| Set-Reset Latch1 | 4 | 4 | 3 | 4 | For high-speed I/O |
| Set-Reset Latch2 | 7 | 7 | 2 | 2 | NOR-based |
| Set-Reset Latch3 | 4 | 4 | 2 | 2 | NAND-based |
| Current Summation Circuit | 24 | 24 | 9 | 10 | For high-speed I/O |
| TSPC Latch* | 9 | 9 | 2 | 2 | - |
| TX Driver1† | 14 | 14 | - | - | - |
| TX Driver2 | 7 | 7 | 10 | 13 | - |
| TX Driver3* | 33 | 33 | 3 | 3 | - |
| Total | 374 | 374 | 225 | 250 | |

† No benchmark samples for bended routing
* 2 different layout designs were used
** 4 different layout designs were used

## A. Generation Task Result

From inputs of benchmark layouts with missing elements, the fine-tuned models inferred generation of them (Table II, Table III). Figure 9 shows example inference results for generation of dummy fingers, N-wells, and routing metals. After legalization, these inferred elements comply with the design rules.

***Contact Generation:*** The model perfectly generated contacts for all benchmark samples. All result layouts passed DRC and LVS verifications. Interestingly, the fine-tuned model was trained with only 50 samples, showing that the foundation model can easily adapt to an easy task of generating simple patterns with a very small training dataset.

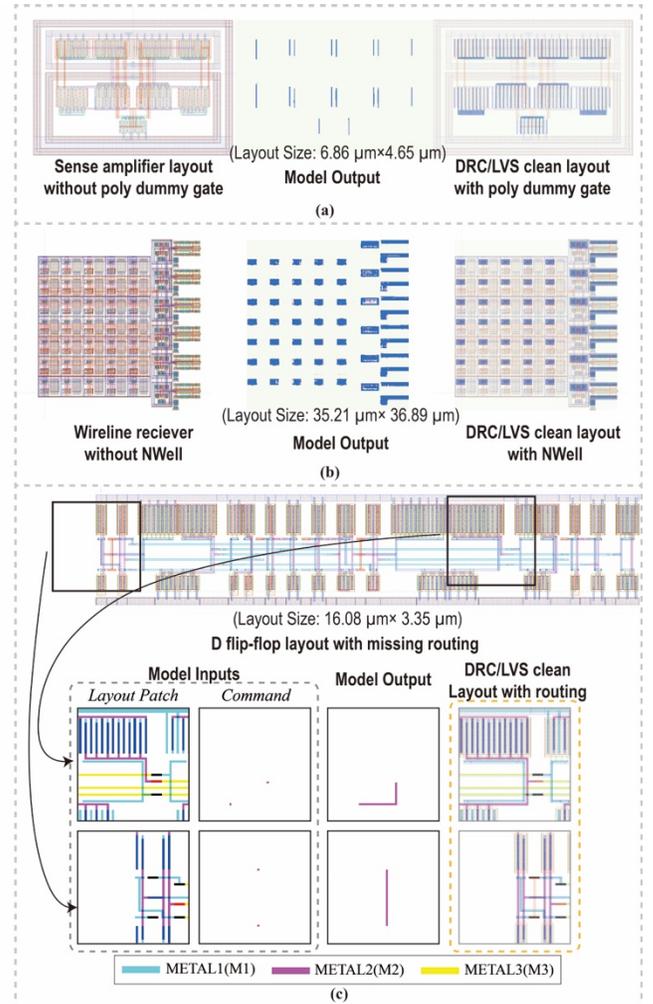

Fig. 9.  Example inference results to generate (a) dummy fingers, (b) N-wells, and (c) routing metals. The layouts of (a), (b), and (c) are shown in different scales.

***Via Generation:*** Vias were generated by trying the top-10 recommendations until the modified layouts passed both DRC/LVS verifications. The model performed particularly well correctly generating all vias (100%) for the largest and most complex circuits (wireline RXs) while the ratio of successful via generation drops for smaller ones, sense-amplifiers (90%) and high-speed set-reset latches (75%) (Table II), resulting in an overall success ratio of 88%. It seems that the contexts of layout patterns of some smaller benchmarks could not provide enough information to infer via generation. This result implies that explicit commands may be necessary to clarify the goal of via generation tasks in order to further improve the success ratio higher than 88%.

***Dummy Finger Generation:*** The fine-tuned model successfully generated dummy fingers for all benchmark samples. Each generated layout passed both DRC and LVS verifications, and the inferred dummy fingers after legalization achieved an IoU score of 0.99 with the original layout designs before dummy fingers removal. Figure 9a shows an example inference result for a strong-arm sense amplifier layout. From the input of the benchmark layout without dummy figures, the model accurately inferred the dummy fingers. After legalization, the result layout successfully passed DRC and LVS verifications.

*N-well Generation:* The model also performed well in generating N-wells. For 298 out of 300 benchmark samples, the result layouts successfully passed DRC/LVS verifications, achieving a success ratio of 99.3%. After legalization, the generated N-wells achieved an impressive IoU score of 0.98. Figure 9b shows an example inference result for a wireline RX benchmark layout with a missing N-well layer. Although the layout is much larger than the patch size, occupying approximately 1000μm2, the model successfully generated N-wells by the proposed slicing and reassembling approach. The result layout passed DRC after legalization.

*Metal Routing Generation:* For the metal routing task, the model successfully generated DRC/LVS-clean routing metals for 599 benchmark samples out of 624 (Table III). Although the foundation model was not pre-trained with explicit commands, the foundation model adapted well when these commands were added during fine-tuning and inference. It is noticeable that the overall success ratio of 96% is much higher than 88% of via generation even though inference on metal routing cannot be easily determined only with the contexts of layout patterns in many benchmark samples. The additional contextual information provided by the explicit commands seems very helpful in improving the inference accuracy.

Fig. 9c shows two example inference results for patches of incomplete D flip-flop benchmark samples with missing routing metals. The model successfully created a straight metal and a bended metal connecting the start and end points. After legalization, both results passed DRC/LVS verifications.

In overall, the fine-tuned models successfully generated LVS/DRC-clean layouts for 1,762 benchmark samples out of 1,824, achieving a high success ratio of 96.6% while conducting the five different downstream tasks. This result implies that the proposed foundation model can adapt to various different layout tasks. Therefore, it can be a consolidated methodology for various layout tasks, reducing enormous human effort to develop many task-specific models.

### B. Comparison with Models Trained from Scratch

To evaluate the efficiency of fine-tuning, we compared benchmark scores obtained by training models from scratch and by fine-tuning the foundation model. Benchmark scores were calculated based on metrics defined in Table I. We trained each model by gradually increasing the sizes of the training datasets. Figure 10 shows benchmark scores for each task versus the training dataset size.

*Contact Generation:* Both models achieved perfect success ratios with only 5 training patches. This result demonstrates that both models are very efficient in identifying and inferring very simple layout patterns.

*Via Generation:* The fine-tuned model achieved a success ratio higher than 70% with only 150 samples. On the other hand, the model trained from scratch required 8 times (1,200 data samples) as many data samples as the fine-tuned model required to achieve a success ratio higher than 70%. With enough data samples (1,500), the fine-tuned model achieved a much higher score than the model trained from scratch.

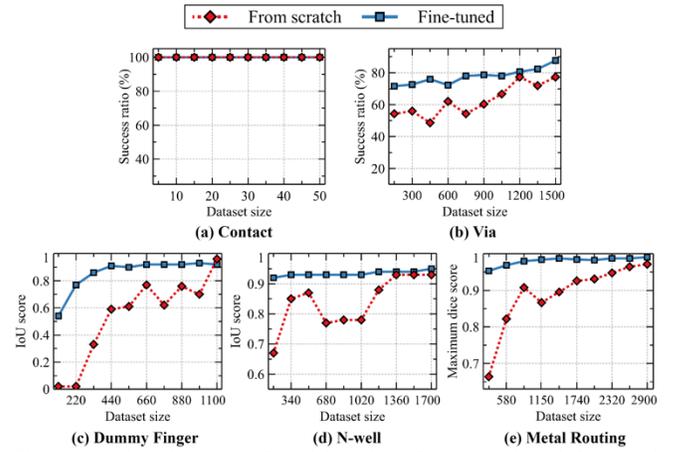

Fig. 10. Benchmark scores of models trained from scratch and fine-tuned from the foundation model versus the dataset size.

*Dummy Finger Generation:* The fine-tuned model achieved an IoU score of 0.54 when trained with only 110 samples. However, as data volume increased to 440 samples, the performance improved significantly, reaching a 0.9 IoU. In contrast, the model trained from scratch scored almost 0 IoU even with about 220 samples. To exceed an IoU score of 0.9, the model trained from scratch required 1,100 data samples, 2.5 times as many data samples as needed by the fine-tuned model. This comparison shows that fine-tuning can achieve high performance with less task-specific data.

*N-well Generation:* The fine-tuned model surpassed an IoU of 0.9 with only 170 samples whereas the model trained from scratch required 1,360 samples (8x) for the same performance. Moreover, the performance of the model trained from scratch significantly fluctuates as the data size increases, potentially due to overfitting on small and biased datasets whereas the foundation model's score increases monotonically with training data.

*Metal Routing Generation:* The fine-tuned model achieved a dice score higher than 0.95 with only 290 samples. To achieve a dice score about 0.95, training from scratch required 2,320 samples, which is 8 times as many samples used in fine-tuning. Furthermore, when the same 290 data samples were used, the fine-tuned model achieved a 40% higher dice score (0.953) than the dice score of 0.664 achieved by the model trained from scratch. These results show the great benefit of the foundation model when only small amounts of training data are available for a downstream task.

In overall, fine-tuning generally outperformed training from scratch, especially with a small training dataset. The foundation model's general knowledge on proper layout patterns enables its efficient adaptation to diverse tasks with small task-specific datasets. Experiments demonstrate that fine-tuning requires much less data than training from scratch to reach the same performance.

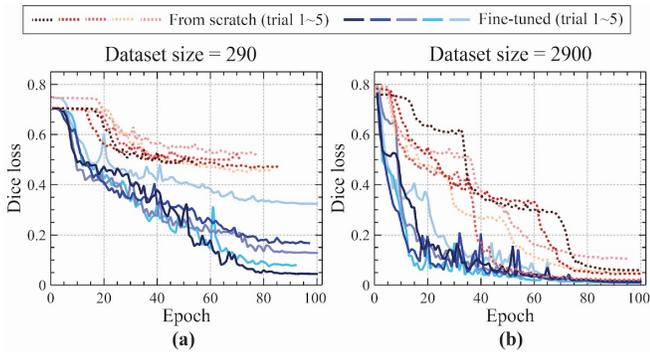

Fig. 11. Validation loss trajectories of models trained from scratch and fine-tuned models. Each model was trained using (a) 290 data samples and (b) 2,900 data samples. Each training process was repeated five times.

### C. Training Convergence Comparison

To compare the training efficiencies of fine-tuning the foundation model and training models from scratch, we trained both models for metal routing tasks using datasets of 290 and 2,900 samples. Each training was repeated five times.

Figure 11a shows the validation loss trajectories of fine-tuning and training from scratch using 290 task-specific data samples. The validation losses of the fine-tuned models monotonically decrease with epoch, reaching a minimum dice loss of approximately 0.045, which is 90% lower than 0.45 achieved by models trained from scratch. These high validation losses were caused by overfitting. The models trained from scratch are likely biased by the small training dataset rather than acquiring knowledge widely applicable to various patterns. This result implies that the fine-tuned models outperform the models trained from scratch if only small amounts of training data are available.

Figure 11b compares the validation loss trajectories at 2,900 data samples, which is 10 times the amount used in the previous experiment shown in Fig. 11a. Both models achieved comparable validation losses below 0.1. However, the fine-tuned models reached a validation loss less than 0.1 in 26 epochs on average whereas the models trained from scratch needed around 70 epochs. This 63% reduction in iterations indicates that even with sufficient data, fine-tuning the foundation model offers significantly faster convergence than training a model from scratch.

These results show that fine-tuned models have a lower risk of overfitting and faster convergence. The pre-trained weights of the foundation model provide a good initial point for optimization of the model's weights for a specific downstream task. Notably, even with the 1/10 learning rate compared with training from scratch, the fine-tuned model converges much faster. This is particularly advantageous for analog layout design automation, where task-specific data are typically rare and large computation resources are usually required.

### IV. Conclusion

We propose the first foundation model that provides a consolidated methodology for diverse analog layout automation tasks and the first self-supervised learning method that addresses lack of labeled analog layout data. The random patch sampling and random masking techniques greatly augmented the data, making them less biased and equally sized to improve the model's generalization ability and to reduce the risk of overfitting during training. The foundation model was implicitly trained for general knowledge on the appropriate layout patterns, and then it was fine-tuned for diverse analog layout tasks. The fine-tuned models successfully suggested the appropriate layout modification such as generating vias, contacts, dummy fingers, N-wells, and metal routings based on the context of layout patterns as well as input commands.

In experiments, the fine-tuned models successfully created proper layouts for 96.6% of 1,824 benchmark samples while conducting the five different downstream tasks. This high performance implies that the foundation model can easily adapt to various layout design tasks, reducing the human effort to explicitly build a separate model for each task.

Comparative analyses show that the fine-tuned models outperformed models trained from scratch, especially with small task-specific training datasets. Fine-tuning also has a lower risk of overfitting than training from scratch because good initial weights are provided by the pre-trained foundation model. Additionally, fine-tuned models converge faster and require less data to achieve the same or better results.

Although we have not explored many similar other tasks and model variants yet, the experimental results demonstrated that our approach can be a key enabler to develop a consolidated methodology for future analog layout automation. The proposed self-supervised learning method resolves the critical challenge of insufficient labeled analog layout data. By pre-training the foundation model for the general knowledge on proper layout patterns and fine-tuning it for various analog layout design tasks, we can resolve diverse challenges in analog layout design with the proposed approach.

### Acknowledgment

The EDA tool was supported by the IC Design Education Center (IDEC), Korea.

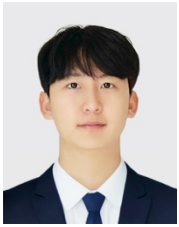

**Sungyu Jeong** (Graduate Student Member, IEEE) received the B.S degree in electrical engineering from Pohang University of Science and Technology (POSTECH), Pohang, Korea, in 2018, where he is currently pursuing the Ph.D. degree. His current research interests include the layout automation of analog circuits, as well as design automation enhanced by deep learning.

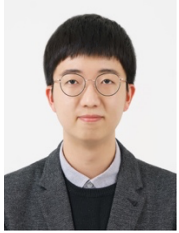

**Won joon Choi** (Graduate Student Member, IEEE) received the B.S. degree in electronic and electrical engineering from Pohang University of Science and Technology (POSTECH), Pohang, Korea, in 2022, where he is currently pursuing the Ph.D. degree. His research interests include high-speed link circuit and computer aided design.

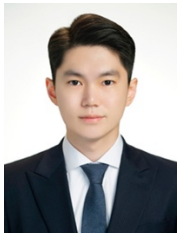

**Junung Choi** (Graduate Student Member, IEEE) received the B.S. degree in electrical and computer engineering from the University of Seoul, Seoul, South Korea, in 2021, and the M.S. degree in electrical engineering from the Pohang University of Science and Technology (POSTECH), Pohang, South Korea, in 2023, where he is currently pursuing the Ph.D. degree in electrical engineering. His research interests include high-speed link circuits and analog layout automation.

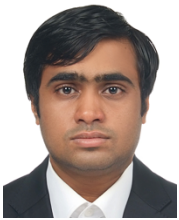

**Anik Biswas** (Graduate Student Member, IEEE) received the B.Tech. degree in Electronics and Communication Engineering from the National Institute of Technology (NIT), Durgapur, India in 2021. He is currently pursuing the M.S. degree in Electrical Engineering at the Pohang University of Science and Technology (POSTECH), Pohang, South Korea. His current research interests include high-speed serial links and design automation.

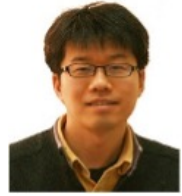

**Byungsub Kim** (Senior Member, IEEE) received the B.S. degree in electrical engineering from the Pohang University of Science and Technology (POSTECH), Pohang, South Korea, in 2000, and the M.S. and Ph.D. degrees in electrical engineering and computer science from Massachusetts Institute of Technology (MIT), Cambridge, MA, USA, in 2004 and 2010, respectively. He was an Analog Design Engineer with Intel Corporation, Hillsboro, OR, USA, from 2010 to 2011. In 2012, he joined the faculty of Department of Electrical Engineering, POSTECH, where he is currently a professor.

Dr. Kim received several honorable awards. He received the IEEE Journal of Solid-State Circuits Best Paper Award in 2009. In 2009, he was an also co-recipient of the Beatrice Winner Award for Editorial Excellence at the 2009 IEEE International Solid-State Circuits Conference. For several years, he served as a member of Technical Program Committee of the IEEE International Solid-State Circuits Conference and has been serving as the Chair of Wireline Sub-com of the IEEE Asian Solid-State Circuit Conference.